# Not a Swiss Army Knife: Academics' Perceptions of Trade-Offs around Generative AI Use

**Razi, Afsaneh** Drexel University, USA | afsaneh.razi@drexel.edu
**Bouzoubaa, Layla** Drexel University, USA | lb3338@drexel.edu
**Pessianzadeh, Aria** Drexel University, USA | ap3943@drexel.edu
**Seberger, John S.** Drexel University, USA | john.s.seberger@drexel.edu
**Rezapour, Rezvaneh (Shadi)** Drexel University, USA | shadi.rezapour@drexel.edu

## ABSTRACT

Our goal is to advance our empirical understanding of the direct engagement of knowledge workers in academia with generative AI (Gen AI), as they are the thought leaders in our society. We interviewed 17 knowledge workers, including faculty and students, to investigate the social and technical dimensions of Gen AI from their perspective. Knowledge workers expressed worries about Gen AI undermining trust in the relationship between instructor and student and discussed potential solutions, such as pedagogy readiness, to mitigate them. Additionally, participants recognized Gen AI's potential to democratize knowledge by accelerating the learning process and act as an accessible research assistant. However, there were also concerns about potential social and power imbalances stemming from unequal access to such technologies. Our study offers insights into the concerns and hopes of knowledge workers about the ethical use of Gen AI in educational settings and beyond, with implications for navigating this new landscape.

## INTRODUCTION

Remarkable effort spanning the computing domains (e.g., natural language processing [NLP], human-computer interaction [HCI], AI, etc.) is being devoted to understanding the sociotechnical implications of Generative AI (Gen AI) technologies (e.g., Baldassarre et al., 2023; Skjuve et al., 2023). In particular, the deployment of ChatGPT has led to extensive research into the positive and negative implications in education (e.g., Busker 2023; J. Yang, 2023).

There is relatively little research directly addressing Gen AI in the context of "knowledge work" (cf L. Li, 2023) – or labor involving the creation, dissemination, and management of information (Slota et al., 2023). Recent studies have identified the need for specific frameworks around academic integrity (Smolansky et al., 2023; Vargas-Murillo et al., 2023; J. Yang, 2023) to mitigate harmful content and negative social impact. Yet, a gap remains in understanding the nuanced implications of these models within academic knowledge work itself. The integration of Gen AI tools into knowledge-intensive professions, spanning writers, researchers, and students, warrants further exploration into the ethical trade-offs faced by people in academia. This work addresses this gap by answering the question: **RQ**: *How do knowledge workers in academia perceive of trade-offs involved in the use of Generative AI?*

To answer our RQ, we conducted semi-structured interviews with 17 knowledge workers professionally situated in academia (i.e., faculty, students, teachers, and librarians). Thematic analysis (TA) (Braun and Clarke, 2012) was used to produce systematic and situated knowledge about our participants' perceptions of Gen AI in relation to their knowledge work roles. We found six prominent themes around general and knowledge work-specific challenges. We discuss these themes and highlight the ethical considerations arising from the opportunities and challenges presented by Gen AI in academia. Concerns about model training, the use of Gen AI as a writing assistant, growing distrust among students and educators, and the need for appropriate pedagogy are reported, illustrating participants' concerns and hopes regarding how Gen AI is reshaping workflows, and redefining the nature of knowledge work. Findings revealed promise and uncertainty: while Gen AI may augment human capabilities and democratize information, it raises concerns about potential biases in AI outputs and the erosion of critical thinking skills.

Our work contributes to the important research areas of ethics, trust, and responsible development in the application of Gen AI tools within academic knowledge work. By addressing these key areas, we contribute to the ongoing discourse on designing AI systems that are not only technologically advanced but also ethically sound and socially responsible in educational contexts. Our research advocates for a balanced approach that considers both technological innovation and ethical implications, aiming to inform the development of AI-enhanced educational tools and environments. Our findings can help shape the integration of Gen AI in academic settings, ensuring that these technologies benefit the educational community, respect individual rights, and promote equitable learning.

## RELATED WORK

### Gen AI Broader Challenges

#### *Accessibility, Governance, and Regulations*

The emergence and widespread availability of advanced AI tools like OpenAI's ChatGPT have highlighted the importance of equitable access to AI technologies. Fairness in accessing AI tools, democratization of AI models (Pinto et al., 2023), and the facilitation of access for people with less technical skills and knowledge (Murgia et al., 2023) have become more prominent (Teixeira et al., 2022). The digital divide, initially described as the disparity in access to the internet (Knowles et al., 2023; Van Dijk, 2006), has evolved to include a second-level division concerning differences in internet skills and usage across various groups and cultures (Scheerder et al., 2017). Gen AI technologies potentially exacerbate this divide. Factors like lack of device access, internet restrictions, and cultural barriers highlight the need for more accessible AI and AI literacy training (Bozkurt and Sharma, 2023; Dwivedi et al., 2023). Gen AI may increase income inequality and market monopolization (Zarifhonarvar, 2023).

Algorithmic bias and ethical considerations in AI development (Faruk et al., 2023; Khowaja et al., 2023) can impact marginalized communities (Knowles et al., 2023). Regulations and clear guidelines for AI governance are crucial to ensure responsible AI deployment and minimize potential harm (Gadiraju et al., 2023; Teixeira et al., 2022, Choung et al., 2023). There is a growing body of research dedicated to examining ethical implications, regulatory requirements, transparency, and governance aspects of these tools (Baldassarre et al., 2023; Khowaja et al., 2023; Panigutti et al., 2023). Hacker et al. (2023) argue that regulation should prioritize transparency and risk management and broaden content moderation guidelines to include governance in Gen AI models. Pistilli et al. (2023) discuss the interconnectedness of ethics, law, and computer science in influencing the development and implementation of AI systems, highlighting the potential for improved and responsible development and governance through the integration of complementary theories and frameworks from these disciplines. Previous research underscores the critical importance of prioritizing transparency, explainability, and interpretability in the future development of Gen AI tools to mitigate potential harms, including the spread of misinformation, risk assessment, liability, fact-checking, regulatory requirements, and the enhancement of AI literacy (Teixeira et al., 2022; Zhan et al., 2023). These accessibility and governance issues have implications for educational technology; therefore, our study adds to the literature, particularly in terms of academic perceptions of these tools in learning systems.

#### *Data Accuracy and Privacy*

LLMs, characterized by their billions of parameters, require substantial training data (Pinto et al., 2023). This data, frequently collected from online platforms, often without explicit consent from the users or creators of the content, includes sensitive and private user information and is subject to inherent biases (Bender et al., 2021). When LLMs are trained on public online data, there is a risk of perpetuating these biases, which can result in a range of harms and challenges for users interacting with these models (Zhan et al., 2023). On one hand, the collection of more extensive data can lead to improvements in the model's performance, enhancing its accuracy and effectiveness and on the other hand, there is a need to prioritize user privacy and ethical considerations (Q. Yang, 2021). This dilemma poses a challenge and necessitates careful research to better understand the trade-offs and balances required in ethical training and developing these models (Baldassarre et al., 2023). As shown in previous research, privacy concerns were identified as one of the major societal implications of Gen AI tools (Baldassarre et al., 2023). In addition, manifestations of false authorship and plagiarism, hallucinations, and accidental spread of misinformation are highlighted as other harmful aspects of ChatGPT (Baldassarre et al., 2023; Ji et al., 2023; Z. Li, 2023). Hallucination, defined as the generation of a piece of text by LLMs based on their internal synthesizing algorithms, rather than the true context or real evidence (Ji et al., 2023), can be especially problematic in sensitive areas such as healthcare or the legal field (Z. Li, 2023). Researchers analyzed the nature and dangers of misinformation created by AI and identified linguistic variances between human and AI-generated content (Gadiraju et al., 2023).

### Gen AI in Education: Trust, Ethics, and Adaptation

While Gen AI could transform communication, it has sparked concern and speculation about its impact on various professions, both in education and beyond (Inie et al., 2023; McKee and Porter, 2020; Taecharungroj, 2023). A key concern with Gen AI is the "death of the author," signaling a shift in how we create and perceive writing, raising doubts about originality, authorship, and the reliability of AI-created content. Some research shows ChatGPT aids creative writing by helping generate ideas, overcome writer's block, and provide feedback, benefiting learners in the process (Mosaiyebzadeh et al., 2023), while other studies have raised concerns. For instance, Inie et al. (2023) point out that professionals are worried about the widespread use of advanced technologies like ChatGPT, fearing they could lead to a decline in quality, harm the creative process, and pose copyright issues. Concerning the adoption of Gen AI in higher education, Yang (2023) analyzed public opinion on a Chinese platform and found data security and student cheating were key concerns about ChatGPT in education, underscoring the need for policies to curb academic misconduct.

Furthermore, the potential use of ChatGPT in teaching and learning has been explored in computer programming (Biswas, 2023; E. Chen et al., 2023), data science education (Zheng, 2023), economic and finance (Lee, 2023). In a thematic analysis, Shoufan (2023) identified 8 positive (helpfulness, ease of use, and human-like conversation) and 7 negative (inaccurate answers and potential malicious misuse) themes about ChatGPT. J. Yang (2023) found ChatGPT and similar tools helpful for self-paced, personalized learning and for easing educators' workloads by handling repetitive tasks. Similar studies found LLMs useful for creating educational content, improving student engagement, personalizing learning experiences, and functioning as a tutor (Kasneci et al., 2023; Rajala et al., 2023). On the other hand, Mosaiyebzadeh (2023) argued ChatGPT can pose different challenges, such as the potential for cheating on exams and homework, risking academic integrity. Previous work identified some of the threats and ethical challenges posed by this technology that need to be addressed, including data privacy and security, transparency, disclosure, lack of fact-checking, accidental plagiarism, and absence of emotions in teaching (Abbas, 2023; Mosaiyebzadeh et al., 2023). Other work raised concern about ChatGPT's potential to hinder critical thinking and other essential skills in students (Smolansky et al., 2023; Vargas-Murillo et al., 2023). These studies emphasized the critical need for specific guidelines, increased classroom assessments, and mandatory reporting of AI-based technology usage. Additionally, they argue for the implementation of appropriate regulations and ethical frameworks to maximize the educational benefits of ChatGPT while minimizing plagiarism and other academic concerns (Baldassarre et al., 2023; Rajabi et al., 2023). However, this literature did not focus on academics in highly technical fields and their perspective on Gen AI in the early stages of releasing ChatGPT, which was widely adopted among knowledge workers, which we will provide more insight on.

## METHOD

### Study Design, Rationale, and Data Collection

We designed a semi-structured interview protocol to understand how knowledge workers perceive the ethical trade-offs related to the use of Gen AI, focusing on their interactions with ChatGPT and similar tools. We engaged with academic knowledge workers (e.g., students and faculty). Our decision to focus on this group was twofold: (1) they represent a key demographic directly using and impacted by Gen AI (e.g., for activities like writing or translating); and (2) we believe individuals active in academia would have the technical background to offer nuanced perspectives on the ethical considerations and potential for human-AI collaboration with Gen AI in their fields.

The semi-structured interviews were divided into three sections. The first section focused on understanding participant familiarity with ChatGPT (e.g., "How did you become familiar with ChatGPT?"). Contextual information included in responses to questions from the first section situated responses to the second interview section: eight questions intended to elicit information about how participants perceive Gen AI in general (e.g., "How do you think ChatGPT can impact people's lives?"). The third section of the interview included seven specific questions about participants' use of ChatGPT in knowledge work specifically (e.g., "What are some scenarios in which ChatGPT can be used in academic research and teaching?", "Do you have any concerns about how these systems could be used in academia?", "What are some possible solutions for these concerns and challenges? If you can think of any." and "Do you think that with our current knowledge and capabilities, we are ready for tools such as ChatGPT?"). Conducted via Zoom, interviews ranged from 40 to 70 minutes, with a mean of roughly 50 minutes. Participants received a $15 Amazon gift card as compensation for their time and effort. Interviews were conducted between April and May of 2023, roughly five or six months following the public deployment of ChatGPT.

| Participant # | Highest Degree | Occupation | Field | Gender | Ethnicity |
|---|---|---|---|---|---|
| P1 | Master's Degree | Student | Civil Engineering | Male | African American |
| P2 | Bachelor's Degree | Student | Information Science | Male | Other |
| P3 | Master's Degree | Faculty | Education | Male | Caucasian |
| P4 | Master's Degree | Student | Computer Science | Female | Caucasian |
| P5 | Doctorate | Faculty | Computer Science | Male | Caucasian |
| P6 | Doctorate | postdoc | Information Science | Female | Caucasian |
| P7 | Master's Degree | Academic Librarian | Digital Initiatives Libraries | Female | Caucasian |
| P8 | Master's Degree | Faculty | New Media | Male | Caucasian |
| P9 | Bachelor's Degree | Student | Computing and Information Sciences, Environment, and Policy | Male | Asian |
| P10 | Master's Degree | Student | Economics | Male | Caucasian |
| P11 | High school graduate | Student | Data Science | Female | Asian |
| P12 | Master's Degree | Student | Computer Science | Male | Asian |
| P13 | Doctorate | Faculty | Information Science | Female | Asian |
| P14 | Bachelor's Degree | Student | Computer Science | Male | Asian |
| P15 | Doctorate | Faculty | Computing and Information Sciences | Female | Middle eastern |
| P16 | Master's Degree | Teacher | Computer Science | Male | Caucasian |
| P17 | Doctorate | Faculty | Comparative Cultural Studies | Self-identify-Genderqueer | Caucasian |

**Table 1. Participants Demographics**

### *Participant Recruitment & Demographics*

We interviewed 17 participants recruited through purposive snowball sampling via social media. We continued data collection until we reached data saturation, which was enough data to address the goal of the study (Malterud et al., 2016). Our number of participants is also adequate given the qualitative nature of the study and aligns with the mean of interview participants from published interview studies being 16 participants (Caine, 2016). Participant demographics are depicted in Table 1. All participants were knowledge workers in US-based academic institutions, including six identified as faculty, eight as students, and three as postdoctoral, librarian, and teacher. Ten participants self-identified as male, six as female, and one as non-binary.

### *Ethical Considerations*

We obtained IRB approval from our Institution, which deemed this study exempt. Prior to participation, all potential participants were provided with a Study Information sheet to review, and their agreement to participate was obtained through an online survey. We took extensive measures to safeguard participant privacy. Interview transcripts were anonymized before data analysis, and we ensured that all quotes presented in our manuscript were similarly anonymized. We were mindful of potential impacts on participants' academic work and relationships, structuring our questions and interactions to minimize any negative effects. All authors of this paper have used Gen AI models and have backgrounds in information and computer sciences which has informed our approach.

## Data Analysis

We employed thematic analysis (TA) (Braun and Clarke, 2012) to analyze our data, with a particular focus on identifying insights relevant to the implementation of AI-enhanced educational tools and learning environments. We first anonymized and transcribed the interviews. Two research assistants (RAs) conducted the initial open-coding of data to develop a broad range of potential codes. During the open coding phase, RAs regularly met with senior members of the research team to discuss and refine the evolving codebook, ensuring that it captured the nuances of Gen AI use in educational contexts. The two RAs independently coded 10% of the interviews using the finalized codebook. We used Cohen's Kappa to calculate the agreement between the two coders (McHugh, 2012) and reached a sufficient agreement ($\kappa > 0.6$). The RAs then coded the remaining interview data. After the coding was complete, the research team met to group the code conceptually into themes. The final codebook, final themes, and the prevalence of such themes are displayed in Table 2. In the following section, we narrate our findings, dedicating one subsection to each of the six final themes presented in Table 2, highlighting their implications for the integration of Gen AI in academic and educational settings.

| Theme | Code | N | % | Definition |
|---|---|---|---|---|
| **Challenges in Data Reliability, Accuracy, and Governance** | Data Training issues | 17 | 100 | The data ChatGPT was trained on may not be reliable and there are possible biases and copyright infringements. |
| | Inaccurate Info | 15 | 88 | Gen AI has the potential to generate inaccurate information, posing a challenge for users to discern its accuracy. |
| | Personal Data | 11 | 65 | Concerns about data governance, data privacy and use. |
| **Human-AI Collaboratio n in Writing** | Writing Tool | 11 | 65 | Gen AI assisting with formatting citations, proofreading, and citation creation. |
| | Research Assistant | 7 | 41 | Facilitating efficient research tasks like literature reviews and data analysis. |
| | Lack of Voice | 9 | 53 | Gen AI leading to a loss of individual writing voice and homogenization of essays. |
| | Writer's Block | 4 | 24 | Overcoming writer's block and guiding paper structure by offering prompts and suggestions. |
| | Overreliance | 5 | 29 | Gen AI encouraging excessive reliance on Gen AI to the point where it negatively impacts the user's critical thinking. |
| | Ideation | 2 | 12 | Gen AI assisting in research brainstorming and question formulation. |
| | Lack of Creativity | 3 | 18 | Excessive use of AI to generate original concepts may lead to concerns about AI's role in the decline of creative process. |
| **Trust and Ethics in Education** | AI Detection Services | 12 | 71 | Exposing the lack of reliable AI detection software and transparency in detection methods. |
| | Cheating | 10 | 59 | Gen AI creates potential for academic dishonesty through plagiarized assignments. |
| | Creating Distrust | 7 | 41 | Teachers using AI as plagiarism detection leading to suspicion and distrust. |
| | Proper Use | 7 | 41 | Gen AI sparks debate on what tasks it is ethical and unethical to delegate to it. |
| **Adapting to AI in** | Pedagogy Readiness | 16 | 94 | Underscores the need for educators to adapt to an AI-infused learning environment. |
| | Tutor | 7 | 41 | Gen AI serving as a personal tutor to answer student questions 24/7. |

| **Education: Challenges and Opportunities** | Faster Education | 6 | 35 | Gen AI promotes faster education by increasing learning pace, covering more material, and raising expectations. |
|---|---|---|---|---|
| | Old-School Education | 4 | 24 | Shift toward traditional learning methods in response to AI concerns, questioning the long-term impact on pedagogical approaches. |
| | Advanced Courses Unpreparedness | 2 | 12 | Exposing the reliance on AI in lower-level courses leading to difficulties in higher-level studies due to lack of fundamental understanding. |
| **Bridging Knowledge & Social Class Gaps** | Socioeconomic Class | 10 | 59 | Gen AI's impact on social classes. |
| | Democratization | 4 | 24 | Gen AI can provide more access to knowledge and the desire to learn for everyone. |
| | Inclusion | 2 | 12 | AI can assist people with disabilities or health conditions. |
| **Practical Solutions from AI Literacy to Regulations** | Educate People | 12 | 71 | Need for education for faculty and students to learn about Gen AI and its limitations. |
| | Call for Regulation | 10 | 59 | Need for AI regulations and new policies. |
| | Built-in Ethics | 7 | 41 | Need to build in ethics to reduce harms. |
| | Development Pause | 1 | 6 | Pausing the development of AI to address the safety measures. |

**Table 2. Codebook and Themes**

## FINDINGS

### Challenges in Data Reliability, Accuracy, and Governance

All seventeen participants expressed concerns about the training data used for Gen AI. Such concerns included: bias baked into Gen AI by way of training data, breach of copyright law in the training of models through unauthorized use of copyright material, outdated information, and lack of fact-checking. When discussing such awareness, participants recommended thorough oversight of model training. Yet recommendations for oversight dissolve into the complexity of factors associated with participant perceptions about Gen AI. Participants questioned the credibility of ChatGPT's data sources and emphasized the importance of establishing mechanisms to improve reliability and accuracy of information generated by these models: *"Where is ChatGPT getting the information? But what exactly is the evaluation process for this information? How I will show that this information is valid and they are quality information. So if we are not sure about these things, and maybe by mistake or by some silly algorithmic function, it pulls out some inferior information, if I may use the word, and you just fly with that information, then we might be facing another pandemic of misinformation"*- P2

Participants expressed concerns about biases arising from training data, such as racial and gender biases. Participants also raised issues about general cultural insensitivity. In short, participants expressed opinions summarized by the following quote from P14: "*the machine learning model is only going to perform as well as the data is based on*." Such concerns are further complicated by the opacity of models. For example, P7 worried that developers of Gen AI systems do not clearly understand how models generate outputs. Such opacity makes it hard to verify the semantic content of language produced by ChatGPT: *"I'm very concerned about the potential for this kind of AI, and the biases that could be in it because it is so opaque, like where it's pulling information from how it's generating that information is completely opaque, that is, up to the developers to design. And my understanding is that in many cases, they don't really understand how it's coming up with the results from the datasets they feed in."*-P7

Such concerns were entangled with questions about privacy concerns and the copyright and ownership of the generated content. Participants further noted concern about models trained on outdated data, which cannot capture timely and local community events.

Fifteen participants (88%) expressed concerns about AI "hallucinations" and potential inaccuracies and false information generated by AI and their lack of understanding regarding how Gen AI chatbots generate content: *"It's not a Swiss army knife where you can just use it in every situation. You need to be very precise with what you are using it for, and not knowing what its limitations are. [...] you cannot use it to verify facts, because it doesn't know what is, it doesn't have it doesn't give you references of any sort. People should be made aware of the hallucinations that it does, because it sounds very confident when it gives answers and if somebody doesn't know that, hey, it could be wrong, then they are going to suffer the consequences of that, and that is going to be the biggest negative impact in my opinion."*- P12

Participants further expressed concerns about Gen AI sources of information and the lack of fact-checking mechanisms. The fear is that users might blindly trust the information provided without verifying it, leading to the spread of misinformation: *"I think a lot of people believe what it tells them, and if there are no safeguards, it could cause a lot of misinformation. And I think misinformation tends to snowball, so I don't know what guardrails exist on ChatGPT on where it's pulling its information from. [...] Fact-checking is also a big mountain to climb, so I don't know if the Internet can truly ever be fully fact-checked. […] So I think it's harmful for people who believe anything that technology tells them without doing their own research."* -P9

Eleven participants (65%) expressed concerns about data governance, data privacy, and use. They argued that the public should have more insight and control over the data being used by ChatGPT. Even the participants who were not as worried about data privacy and governance compared with their social circles also conveyed concerns: *" It's worrying that the company won't tell us how long my queries are held. How identifiable are they who have access to what I have searched for? And when would that happen?"*-P3 These concerns about data reliability and accuracy are particularly relevant in educational contexts, where the integrity of information is crucial for effective learning outcomes.

### Human-AI Collaboration in Writing

Knowledge workers involved in research have recognized the value of Gen AI as a writing tool. They mentioned its potential to assist with proofreading for grammatical errors, and formatting citations (N=11, 65%). One participant even compared ChatGPT to asking their spouse for proofreading assistance, highlighting the convenience it offers. Additionally, participants acknowledged the potential of Gen AI as a research assistant (N=7, 41%) during the initial phases of research, particularly for ideation. They have mentioned its current use for ideation and the possibility of enhancing the research ideation process by providing real-time insights on the current state of the art and identifying research gaps - a feature not currently available in general tools like ChatGPT due to gaps in training data: *"I would say, I do think just in terms of the process of developing a research project. I mean, my experiences with it are that accepting the limitations that exist can be very useful, almost as a research partner, right? Because it does have access to this massive dataset that it can pull up instantly. And I think there are a lot of potentials to develop and refine ideas and projects and things like that and engagement"*-P7

Users mentioned that Gen AI can be a helpful tool for addressing "writer's block" (N=4, 24%) as well as assisting with grammatical errors, formatting citations, and general ideation. They claimed that ChatGPT makes it easier to write by providing guidance for their paper, preventing them from spending excessive time staring at a blank page, and allowing them to write more efficiently.

Furthermore, participants (N=9, 53%) expressed dissatisfaction with ChatGPT's inability to capture personal voice and style, particularly in tasks like writing cover letters or essays. They noted that while ChatGPT produces acceptable content, it often lacks the unique personal touch that makes such communications authentic: *"...But I still think it does the job as in like it gives you a cover letter that's acceptable, but something that I did not like about it was, and it's pretty obvious, it doesn't feel like me, and I think when we need something like a cover letter, or like you're writing an essay for a class, it's personally important for me to feel like this is my voice."* -P9

The issue of potential overreliance (N=5, 29%) on ChatGPT was also raised. Participants (N=3, 18%) worried that the excessive use for assignments could hinder critical thinking and diminish the benefits of collaborative learning: "*Maybe some people might get so dependent on it. I know some people don't do any of the assignments and only depend on ChatGPT... I remember we had this course, it was very tough, and we used to get together a few people in a group and then talk about it, discussing, Googling it, reading books. But when you use ChatGPT you're not doing any of this. You're just getting some prepared answers. So I think ChatGPT won't let them [people] think."* -P4

Considering the potential consequences of overreliance on Gen AI participants acknowledged the tradeoff of speed vs. authenticity. Using Gen AI to create content may speed up efficiency; however, it could stifle human originality and expression, leading to homogenization of content. They mentioned assuming that the quality of content developed by Gen AI is suitable, the need for human innovation in those areas could decline, leading to a decline in human creative potential and problem-solving abilities.

### Trust & Ethics in Education

The most expressed concern in academic settings regarding Gen AI was the use of AI detection services, which may produce inaccurate or misleading results (N=12, 71%): *"There was a professor in Texas who heard about these tools [ChatGPT] and wanted to make sure students were not cheating on their final papers[...]I think those are wildly irresponsible to use right now. But what he did was paste their essays into ChatGPT and ask, "Was this written by you?" And of course, it came back and said yes, for like 90% of his class, and then he failed them all because he did not understand what that system was doing."* - P5

Another concern was cheating (N=10, 59%), especially in writing assignments. Participants were concerned that students, under stress from their coursework, may attempt to submit papers generated by AI to meet deadlines instead of asking for extensions or writing their own work: *"They run out of time [for assignments], and that's when they [cheat]. So, we just have to ask questions expecting them to use it [ChatGPT] so that they can't cheat, because that's the expectation."* -P16

Although cheating was frequently expressed, there was also mention of the concern of creating a distrustful culture between students and teachers (N=7, 41%). We observed opposing forces around trust and Gen AI as a threat by faculty versus a tool by students. Students expressed that they felt accused of using ChatGPT to bypass learning and not understanding concepts, such as the below quote by a student: *"Yesterday my friend told me that her English teacher wanted to fail her from the class because she used ChatGPT to write an assignment. Of course, one wouldn't want to complete an assignment, so they can just kind of ask ChatGPT to do it. But then that begs the question. What are you grading them on? Are you grading them on original thought and understanding of the concept? Or are you grading them on how many words they wrote, and whether it's grammatically correct? Instead of AI writing for you, if it writes with you and becomes this writing coach, you don't consider that as plagiarism, right? ... We just have to rethink the way we test students." - P14*

We had diverse perspectives on the matter of distrust from faculty and students, such as the fear among students that they would suffer the consequences. Whereas faculty felt that Gen AI needs to be used more responsibly by students: *"I guess my concern is kind of like a combination of the privacy of students and my ongoing philosophy for a long time has been to trust students. Lately that's been challenged; I'm seeing cases where I can't trust the work I'm getting. That I know is not theirs. So that's kind of a concern, you know privacy and trusting students, and then just wanting students to learn. And I mean, to enjoy learning for the sake of learning, that's something that I always try to really emphasize to my students." - P17*

Among the interviews, 41% (N=7) of the participants emphasized the importance of considering the appropriate scenarios for using Gen AI. The mental model of proper usage encourages people to adapt to a changing learning environment where AI is increasingly present. Participants mentioned that when debating which tasks are ethically acceptable to delegate to AI, they explore issues such as accountability, transparency, and fairness: *"For example, if I need to know how to use a saw properly, I can just say what's the good way to use a saw [...] if you know for a fact that information won't affect your health or your livelihood. I think you can use ChatGPT there, because if it's going to affect your health, I wouldn't be asking ChatGPT here how to treat cancer...I might ask that: Hey? You know what, I have a minor cut, what's going to be the best way to disinfect it, so depending on the severity of something, I think ChatGPT can be used in scenarios like that."*-P12

These concerns highlight the need for careful consideration of ethical issues in the design and implementation of AI-powered educational tools and assessment systems.

### Adapting to AI in Education: Challenges & Opportunities

Participants expressed that Gen AI could assist students in quickly identifying and presenting relevant sources of information for their assignments, saving them time compared to traditional internet or textbook searches. Instructors also recognized the potential of Gen AI to modernize the classroom but acknowledged the importance of "pedagogy readiness" - the reflection on how to incorporate AI technology into teaching practices. 94 percent of participants (N=17) emphasized the significance of pedagogy readiness, suggesting that students should be encouraged to critique Gen AI responses instead of solely relying on them to solve problems. This approach would enable students to develop critical thinking skills typically reserved for research-based courses: *"...this sort of paradigm of assignments where you get ChatGPT to do something, and then critique it. Those are pretty useful. I don't think that's the only use case, where there's some educational value to be extracted from these systems other than like special topic in a machine learning course." - P5*

Participants recommended modifying the education curriculum to address the potential negative impacts of Gen AI on education. The suggested changes encompass a wide range of improvements, including crafting more engaging and relevant assignments, refining question formats to challenge critical thinking skills, selecting materials that enhance knowledge acquisition, and adopting fair and effective evaluation methods to accurately measure progress.

Among knowledge workers, participants in (N=7, 41%) academia expressed the potential benefits of using Gen AI as a tool to supplement student education. They believed that ChatGPT could facilitate easier questioning, topic discussion, and addressing writing issues, acting as an at-home tutor: *"... When a student is learning something in class, you can use AI to supplement that learning, and give everybody a good tutor, right, like I mentioned before. If everybody is given a really good tutor, everybody's performance becomes really good. It's not just a below average student becomes an above average student, and an average student becomes an excellent student..."* - P14

Participants discussed the potential for Gen AI to accelerate the pace of education (N=6, 35%). They mentioned that AI could provide quick access to information, allowing students to learn at their own pace and enabling educators to cover more material in a shorter amount of time. However, concerns were also raised about the risk of sacrificing depth of understanding and critical thinking skills in favor of speed. Some participants (N=4, 24%) proposed returning to a more traditional style of education for curriculum and assignments. They suggested having oral exams without tools like PowerPoint to ensure students cannot rely on AI.

Some participants (N=2, 12%) expressed concerns about students relying heavily on Gen AI. They believed this reliance could lead to a lack of foundational knowledge in a topic, making students unprepared for advanced courses and ultimately resulting in failure. These insights can inform the design of AI-enhanced learning environments that effectively integrate Gen AI while maintaining pedagogical integrity and fostering critical thinking skills.

### Bridging Knowledge Gaps, Enabling Accessibility, and Social Dynamics Impacts

Our participants raised several benefits and concerns about the potential impacts of using Gen AI, highlighting a range of issues from societal impacts to personal experiences. A significant concern, expressed by half of our participants (N=10, 59%), was the potential of ChatGPT to either widen or bridge social class divides. The fear was that access to, and the use of advanced technologies like ChatGPT might create or exacerbate inequalities, either by providing advantages to those who can access it or by leaving behind those who cannot.

On the contrary, ChatGPT was particularly appreciated by participants (N=4, 24%) for its role in democratizing access to knowledge, enabling a broader range of individuals to engage with information and learning resources: *"Whether it be academic or non-academic, the gap between the Western society and developing could be like India, has virtually closed because of the Internet. So I think ChatGPT can only take that further [...] I think there are a lot of opportunities in it being like an equalizer tool where people from different parts of the world are coming up with the same by-product because they had access to the same tools." -P9*

Furthermore, participants (N=2, 12%) highlighted ChatGPT's role in aiding people with disabilities. One individual with a visual impairment recounted how ChatGPT acted as a reading assistant, interpreting and verbalizing written content that they otherwise found challenging to access: *"This could also be beneficial for people who are visually impaired, allowing them to understand specific parts of an article or content. It would be useful to take pictures of artifacts and have ChatGPT describe what they are." -P6*

These findings underscore the potential of AI-enhanced educational technologies to address issues of accessibility and equity in learning environments.

### Practical Solutions from AI Literacy to Regulations

Participants came up with practical solutions when asked about overcoming the challenges with Gen AI. Many participants (N=12, 71%) mentioned that there needs to be a better understanding of the technology in people's lives to use Gen AI effectively and ethically. A solution to these issues is to educate people on the capabilities of Gen AI and provide awareness of its limitations. For example, P2 stated that people always find a way to use new technologies even if it is banned with punishments, so the solution is to teach them to use them responsibly: *"I'm mentioning information literacy so much because this is what I know about, and it's something I've been reading. You need to let students know that there is no problem in using it. But you have to use this responsibly. It is more about creating awareness, Train people to know how to use it, and know where to use and not to use it."*-P2

Faculty initiated talks and workshops to educate other colleagues and instructors on Gen AI: "So rather than jump to the idea of banning it, my colleagues and I were more interested in, well, are there ways we can harness this for good? Can we make our classes better rather than worse by using this technology in the classroom? So we've held a series of conferences and webinars on the topic."- P8

Further, many participants (N=10, 59%) called for the need for regulations and policies to control some aspects of the use of Gen AI. They provided examples of cases of how those regulations would be helpful in different sectors. They called these policies and regulations to ensure the ethical use of Gen AI and to limit the power that it has: *"For the government, you should create laws, because if the roots or the source of this data is checked, I think it'll help a lot. I mean, they just need to create laws and policies, we'll make sure that the contents of this data are well-sourced, and it might reduce the chances of giving very bad results at the end of the day."*- P1

Participants (N=7, 41%) expressed the lack of ethics built in for Gen AI. They expressed concerns regarding the potential for harmful application of artificial intelligence when ethical considerations are not embedded within its design. However, integrating ethical principles into AI is also a subject of debate. Participants described different scenarios in which malicious users could use Gen AI to commit crimes and harm others, from stealing from banks (P1) to building bombs (P12). They mentioned many of the safeguards implemented to avoid receiving harmful information could be bypassed by role-playing and jailbreaking techniques: *"Yesterday in class for the first time, students reported that ChatGPT was breaking rules, being asked to roleplay and the reason is that you can get it to do things that are outside of the safety protocols by asking it to take on a role that is somehow like imagining that it's some creature that isn't bound by the rules that it's normally bound by."*-P8

P8 elaborated further that they constantly try to safeguard their models against these attacks, and it is hard to cover all the possible ways that these models could be exploited. Only P8 suggested pausing the development of Gen AI technologies as a potential solution to some of the challenges discussed. The fact that only one participant suggested pausing Gen AI development indicates this was not viewed as a viable solution by most participants. Instead, the majority proposed alternative approaches to address the challenges posed by Gen AI in academia. While public discourse may include calls to pause AI development, the academics in this sample generally believed that pausing the development and use of Gen AI was neither feasible nor desirable and instead offered a range of solutions to promote its responsible use. These proposed solutions have implications for the development of AI literacy curricula, the design of ethical AI-enhanced educational tools, and the formulation of appropriate policies.

## DISCUSSION

### Utility vs. Transparency Trade-offs in Academia

Our findings highlight the tension between the potential benefits of Gen AI and the ethical concerns raised by knowledge workers. While recognizing its potential to improve the efficiency of academic pursuits, participants also expressed concerns about the trade-offs associated with this technology. These include the risk of introducing learned biases, the potential loss of individualism in voice, and over-reliance on Gen AI, which could impact learners' understanding of foundational concepts in the long term. These trade-offs have implications for the design and implementation of AI-enhanced educational tools, particularly in terms of ensuring transparency and maintaining academic integrity. In this section, we will delve into these considerations and their implications among knowledge workers. One concern raised by the participants was the potential misuse of Gen AI as a writing assistant. They expressed worries about academic integrity and trust breaches, whether by educators, students, or research faculty. Already, there has been growing concern about student cheating even prior to the widespread use of Gen AI (Bashford, 2022; Malesky et al., 2022; Stokel-Walker, 2022). Now, students face false accusations from AI plagiarism detectors, leading to a loss of trust in student work by their educators. The limitations of AI detection tools further compound these concerns, as highlighted by Gorichanaz (2023). Our finding that faculty see ChatGPT as a threat, whereas students see it as a tool, implies fundamental, but implicit, misunderstandings about how the educational context (i.e., the social purpose of education) may degrade trust in academic settings.

Another concern raised by participants related to the biases present within the training data used for Gen AI. These models learn from diverse public datasets over the internet, which often reflect societal biases, including racial, gender (Busker, 2023), and disability biases (Gadiraju et al., 2023). As a result, generated content may inadvertently perpetuate and amplify these biases, raising ethical questions about the responsible use of such technologies. The human biases inherent in training data would be magnified in Gen AI systems (Bender et al., 2021), as they will be adopted vastly in different fields across the globe. It is worth noting that one of the major players in this field, such as OpenAI, has reported that a large portion of their training data came from Reddit, a pseudo-anonymous social media platform that is known for its diversity in topics, plethora of communities, and levels of engagement. Such data are mostly in English and produced by WEIRD (Western, Educated, Industrialized, Rich, Democratic) community (Henrich et al., 2010). Also, while this content may be considered more 'high-quality' due to the selection criteria of data (at least 3 karma; Solaiman et al., 2019), it often provokes intense reactions from users and tends to be more controversial. This concept, known as digital emotion contagion (Goldenberg and Gross, 2020), suggests that the bias from emotionally intense training data could impact the Gen AI model's output. Therefore, this warrants more research, especially with communities impacted most by those biases, to help co-design guidelines and ways to address and mitigate biases in training data, which are essential to ensure fair and unbiased outcomes. Best practices on transparent dataset documentation such as Data Card (Gebru et al., 2021; Pushkarna et al., 2022) could be used to create more understanding for various stakeholders of the dataset origins and evolutions for more responsible AI development and deployment. More specific guidelines are needed for Gen AI dataset transparency as they require a larger scale for training the datasets, which are mostly publicly available and need explanations of copyright documentation. As companies optimize the fast-paced production of technology and maximize their profits, policies and regulations should be put in place to provide those explanations.

### Pedagogical Impact

Unlike traditional educational technologies that provide static resources (e.g., pre-recorded lectures) or one-size-fits-all solutions (e.g., uniform grading delivery), Gen AI has the potential for dynamic feedback, tailored content, and individualized support. Our study shed light on this unique potential, as participants primarily identified "pedagogical readiness," "educational supplement," and "writing assistant" functions as key opportunities for Gen AI in academic settings. These findings are in line with current literature that explored the use of AI in the classroom (Dakakni and Safa, 2023; Seo et al., 2021). With the advent of Gen AI, educators are introduced to the opportunity to enhance their pedagogy through tackling specific tasks like automated lesson planning, adaptive assessments, and personalized feedback generation. Gen AI assistance in these cases are advertised to free up more time for teachers to dedicate to human-centered tasks, like mentoring. Within the curriculum development processes, Gen AI has the potential to assist educators in developing more personalized learning pathways that cater to student's individualistic needs. For example, an AI tool can recommend personalized reading materials and subsequently practice problems aligned with their learning gaps. Additionally, these tools can offer real-time feedback to students to increase their learning efficiency and the grading of teachers. With the advantage of having just-in-time, personalized support for large-scale settings, and improving the feeling of connection (Seo et al., 2021), Gen AI could ultimately enhance collaboration between people and machines, with the goal of improving educational outcomes set forth by entities like educational oversight committees, teachers, and parents. For non-native speakers, traditional language learning often proved challenging due to limited opportunities for real-time practice and feedback. Gen AI can bridge this gap by providing personalized and interactive dialogue prompts, adapting to individual errors and preferences, and offering context-sensitive suggestions for improvement (Kamalov et al., 2023). This level of engagement can enhance language acquisition, boost confidence in classroom communication, and bridge communication gaps between learners and educators, offering better collaboration and knowledge sharing. Each of these cases allows for Gen AI to act as a virtual, learning scaffold. In times of targeted support, these technologies can provide guided problem-solving and step-by-step explanations but should gradually fade away as students gain confidence in the topic to prevent an overreliance on the tool and, ultimately, a lack of foundational understanding. There has been a body of research (Karsh, 2018; Straub, 2009) on technology adoption in the workplace, suggesting automation creates more labor, as workers must learn how to use, incorporate, troubleshoot, and maintain the new technology.

While participants mentioned that Gen AI has the potential to bridge the educational gap and provide equal access to knowledge, they were worried about the risk of widening the gap due to unequal access to technology. This widening of the "digital divide" depends on various factors such as access to computers and high-speed internet, as well as a lack of comprehensive understanding of Gen AI, particularly among students of color and the schools that serve them. To ensure that Gen AI is fair and equitable as an educational aid, we need to consider providing necessary technology support to low-resource school districts. Additionally, it is important to develop culturally inclusive Gen AI models to avoid exacerbating existing inequities. This requires frameworks for more ethical data collection. Lastly, to ensure the effective implementation of Gen AI tools, we must provide adequate training and support for educators on how to appropriately utilize Gen AI.

### Data and Algorithmic Governance Requires Careful Power Dynamic Considerations

While participants suggested AI regulations, questions regarding who should regulate AI presented a dilemma, including the potential for repression (Feldstein, 2019). Several countries have already begun to develop executive orders and guidelines to regulate AI (Madiega, 2019; The White House, 2023). While regulations are necessary to curb potential misuse, centralized control, especially in the hands of powerful entities like corporations and governments, raises concerns about censorship and manipulation. For example, the governments of more authoritarian regimes may exploit their power to censor technology and the internet for surveillance purposes, infringing upon the rights and freedom of speech of their citizens (Funk et al., 2023). Even in democratic countries, lobbying may skew policies toward private interests, driven by profits and social control (Rainie et al., 2021) at the expense of individual rights. A potential solution could lie in establishing an independent, global entity to oversee responsible AI practices, accounting for the size and needs of different models and companies. Although the participants suggested having regulations and policies to mitigate the negative impacts of Gen AI, despite our best intentions, any action to implement applications, policies, or interventions may have unintended consequences that may lead to harm. For instance, a well-intentioned policy designed to reduce bias in AI may have the unintended effect of unfairly targeting a particular group of people. Therefore, we must exercise caution and thoughtfulness in our approach to ensure that we do not inadvertently cause harm while striving to make progress.

### Limitations and Future Research

Our study focused on knowledge workers in US academia. Data are subject to self-report bias. Findings are limited to our participants and are not generalizable in a positivist sense. Future research should investigate the perceptions and experiences of knowledge workers across diverse cultures, languages, and backgrounds to gain a more comprehensive understanding of the ethical challenges of using Gen AI in education. Our study opens opportunities for future research employing participatory design methods, real-time assessment, or longitudinal studies to directly address the challenges identified while actively involving educational stakeholders in designing solutions. Future studies could explore the long-term impacts of integrating Gen AI into educational curricula, investigating how it affects student learning outcomes, critical thinking skills, and overall educational experiences. Research could also focus on developing and evaluating AI literacy programs for educators and students, ensuring they can effectively and ethically use Gen AI tools in educational settings. Finally, there is a need for interdisciplinary research that combines insights from education, computer science, ethics, and policy to develop comprehensive frameworks for the responsible integration of Gen AI in education.

## CONCLUSION

We provide valuable insights from interviews with academic knowledge workers from technology-related fields about Gen AI's opportunities, ethical challenges, and solutions. Our findings highlighted tensions of knowledge workers using Gen AI as a writing and research assistant while wondering about challenges around transparency, reliability, and inaccuracy of generated content. These findings have implications for the design and implementation of AI-enhanced educational tools and learning environments, particularly in terms of balancing the benefits of Gen AI with the need to maintain academic integrity and foster critical thinking skills. Moreover, knowledge workers are apprehensive about the impact of Gen AI on instructors and students. The potential of Gen AI to democratize knowledge access highlights both opportunities and challenges for creating more equitable educational experiences. Our research contributes to the ongoing discussions on ethical and effective AI-enhanced educational technologies and pedagogical needs in the design of AI systems for education.

## GENERATIVE AI USE

We confirm that we did not use generative AI tools/services to author this submission.

## AUTHOR ATTRIBUTION

Conceptualization – all authors. Data curation – A. Razi, R. Rezapour conducted the interview. Analysis and Investigation – A. Razi, R. Rezapour, L. Bouzoubaa analyzed the interview data. Funding acquisition – A. Razi, R. Rezapour. Supervision – A. Razi, R. Rezapour. Writing – all authors.